# Socially Compatible Control Design of Automated Vehicle in Mixed Traffic

Mehmet Fatih Ozkan and Yao Ma[1]

*Abstract—* In the car-following scenarios, automated vehicles (AVs) usually plan motions without considering the impacts of their actions on the following human drivers. This paper aims to leverage such impacts to plan more efficient and socially desirable AV behaviors in human-AV interactions. Specifically, we introduce a socially compatible control design for the AV that benefits mixed traffic in the car-following scenarios. The proposed design enables the altruistic AV in human-AV interaction by integrating the social value orientation from psychology into its decision-making process. The altruistic AV generates socially desirable behaviors by optimizing both its own reward and courtesy to the following human driver's original plan in the longitudinal motion. The results show that as compared to the egoistic AV, the altruistic AV significantly avoids disrupting the following human driver's initial plan and leads the following human driver to achieve considerably smaller car-following gap distance and time headway. Moreover, we investigated the impacts of the socially compatible control design with different altruism levels of the AV using statistical assessments. The results collectively demonstrate the significant improvement in traffic-level metrics as a result of the AV's altruistic behaviors in human-AV interactions.

## I. INTRODUCTION

Human drivers usually only consider preceding traffic when planning longitudinal motions in the car-following scenarios and neglect the impacts of their actions on the following drivers due to their egoistic nature and limited perception. This may lead to socially unwelcome behaviors that are perceived as dangerous, uncomfortable, or overly defensive by the following human drivers [1]. It may also lead to inefficient road utilizations such as phantom jams and unnecessary stop-and-go [2]. To overcome such drawbacks, we propose designing a socially compatible automated vehicle (AV) control strategy in the mixed traffic that explicitly considers its impacts on the following human drivers through human-AV interactions. This is enabled by advanced sensing [3] and communication technologies [4] of AVs that are generally unavailable for human drivers. Different from most of the existing work on AV cruise control which only considers egoistic goals using preceding traffic information [5-6], by investigating and leveraging the human-AV interactions in mixed traffic, we aim to generate socially desirable behaviors of AV and quantitatively evaluate the benefits of such design to improve traffic efficiency.

M. F. Ozkan and Y. Ma are with the Department of Mechanical Engineering, Texas Tech University (e-mail: mehmet.ozkan@ttu.edu and yao.ma@ttu.edu).

To generate socially desirable behaviors, it is necessary to establish a suitable and generalizable behavior generation framework of AVs that integrates the social factors in human-AV interactions. Some studies define social factors as selfishness and altruism via Social Value Orientation (SVO) [7]. SVO quantifies the degree for AVs to act egoistically or altruistically in human-AV interactions. The egoist AVs make decisions that only benefit their own utility, and the altruistic AVs optimize a social utility that incorporates the benefits of the human drivers as well. Such altruistic AVs have been demonstrated to create socially compatible outcomes in some realistic driving scenarios, such as highway merging ramps [7-8] and unprotected left turns at intersections [9]. No results have been reported on altruistic AVs in mixed car-following scenarios. Considering human drivers will remain to be primary vehicle operators in the foreseeable future, insights into the impacts of altruistic AVs on mixed traffic will be beneficial for practical and theoretical merits.

In order to achieve socially compliant motion control, altruistic AVs first need to understand and predict the behavior of human drivers. Inverse reinforcement learning (IRL) based driver behavior models have been widely used to understand the behavior of human drivers [9-14]. The IRL approach aims to learn an underlying cost function that encodes the driving preferences of the human driver in driving demonstrations. In our previous work, we have demonstrated that the proposed IRL-based driver behavior model can effectively learn and replicate human drivers' driving preferences in the car-following scenarios [13].

This study explores the potential benefits of developing a socially compatible control design for the automated vehicle in the car-following interaction with a following human driver and investigates the impacts of different altruism levels of the automated vehicle on mixed traffic. The main contributions of this study are as follows: 1) altruism is integrated into the decision-making processes of the automated vehicle to achieve socially compatible behaviors in the car-following interactions with the following human driver. 2) the impacts of such a socially compatible control strategy of the automated vehicle on mixed traffic are analyzed considering the automated vehicle's altruism variations toward the human driver. To the best of the author's knowledge, this is the first study on the socially compatible driving strategy of the automated vehicle in the car-following scenario and its impacts on the traffic flow of microscopic-level mixed traffic.

The remainder of this paper is organized as follows. In Section II, the problem formulation in human-AV interaction is introduced. In Section III, the socially compatible behavior planning is formulated. In Section IV, the socially compatible control design in mixed traffic is developed. Section V examines the socially compatible control design's impacts on mixed traffic with numerical simulations in realistic driving scenarios. Section VI concludes with closing remarks.

## II. PROBLEM FORMULATION

In this study, we consider an interactive two-agent system in the car-following scenario where the agents are AV $\mathcal{R}$ and human $\mathcal{H}$. The objective is to develop an altruistic AV that considers the interest of itself and courtesy towards the following human driver in the human-AV interaction. Let $x_i$ and $u_i$ denote the state and control input of the AV $(i = \mathcal{R})$ and the human driver $(i = \mathcal{H})$, respectively. $x^t$ denotes the state of the human-AV interaction at time $t$, where $x = \left(x_\mathcal{R}^T, x_\mathcal{H}^T\right)$ and satisfies the overall system dynamics, as shown below:

$$x^{t+1} = f\left(x^t, u_\mathcal{R}^t, u_\mathcal{H}^t\right) \quad (1)$$

At time $t$, the longitudinal vehicle dynamics of the AV and human driver are represented by the linearized third-order model [15], and an actuation time-lag is considered between the realized longitudinal acceleration and the control input:

$$\dot{d}_i^t = v_{i,P}^t - v_i^t, \ \dot{v}_i^t = a_i^t, \ \dot{a}_i^t = \frac{1}{\rho}\left(u_i^t - a_i^t\right), \ i \in \{\mathcal{R}, \mathcal{H}\} \quad (2)$$

where $d_i$ represents the gap distance of the AV and human driver with respect to their preceding vehicle in traffic; $v_{i,P}$ denotes the longitudinal speed of the AV's and human driver's preceding vehicle; $v_i$ and $a_i$ represent longitudinal speed and acceleration of the AV and human driver, respectively and $\rho$ denotes the actuation time-lag. The system state of the AV and human driver at time $t$ can be defined as $x_i^t = \left[d_i^t, v_i^t, a_i^t\right]^T$ and formulated as:

$$\dot{x}_i^t = A x_i^t + B u_i^t + D v_{i,P}^t \quad (3)$$

where $A = \begin{pmatrix} 0 & -1 & 0 \\ 0 & 0 & 1 \\ 0 & 0 & -1/\rho \end{pmatrix}$, $B = \begin{pmatrix} 0 \\ 0 \\ 1/\rho \end{pmatrix}$ and $D = \begin{pmatrix} 1 \\ 0 \\ 0 \end{pmatrix}$. The linearized third-order model is discretized with a zero-order hold (ZOH). The discretized version of (3) can be stated as $x_i^{t+1} = A' x_i^t + B' u_i^t + D' v_{i,P}^t$, where $A'$, $B'$ and $D'$ are the discretized version of $A$, $B$ and $D$, respectively.

We consider that both the AV and the human driver are rational planners whose goals are choosing actions to maximize their rewards or equivalent to minimize their cost functions over the planning horizon during the vehicle operation. Therefore, we assume that the optimal control problem of the AV and the human driver can be solved by using the Model Predictive Control (MPC) approach over the planning horizon $N$. Let $C_\mathcal{R}$ and $C_\mathcal{H}$ are the cost function of the AV and the human driver over the planning horizon, respectively:

$$C_i\left(x^t, \mathbf{u}_\mathcal{R}, \mathbf{u}_\mathcal{H}\right) = \sum_{k=0}^{N-1} c_i\left(x^{t,k}, u_\mathcal{R}^k, u_\mathcal{H}^k\right), \ i \in \{\mathcal{R}, \mathcal{H}\} \quad (4)$$

where $x^{t,k}$ represents the $(t+k)^{\text{th}}$ predicted system state of the human-AV interaction and $\mathbf{u}_i = \left(u_i^0, u_i^1, \ldots, u_i^{N-1}\right)^T$ defines the sequence of the predicted control inputs of the AV and the human driver, respectively. At every time step $t$, the AV and human driver can generate their optimal vehicle operations by minimizing $C_\mathcal{R}$ and $C_\mathcal{H}$, respectively, and compute their first control inputs $u_\mathcal{R}^{*0}$ and $u_\mathcal{H}^{*0}$, and replan at time $t+1$.

The closed-loop dynamics of the human-AV interaction can be formulated as a game considering the optimization-based state feedback strategy during the interaction. To simplify this game, we assume that the AV and human driver are running a Stackelberg game where the AV is the leader and the human driver is the follower, as expressed in [10-11]. In the traditional two-agent Stackelberg game, the leader chooses an action and the follower computes its best outcome given the leader's action. With this simplification, we further assume that the AV can access $C_\mathcal{H}$ and the human driver only computes the best response to the AV's actions rather than influencing the AV's original plan. This presumption refers that for every control sequence that the AV considers, the AV can compute how the human driver would respond and how much it would cost to the human driver:

$$\begin{aligned}\mathbf{u}_\mathcal{H}^*\left(x^t, \mathbf{u}_\mathcal{R}\right) &= \arg\min_{\mathbf{u}_\mathcal{H}} C_\mathcal{H}\left(x^t, \mathbf{u}_\mathcal{R}, \mathbf{u}_\mathcal{H}\right) \triangleq g\left(x^t, \mathbf{u}_\mathcal{R}, \mathbf{u}_\mathcal{H}\right) \\ C_\mathcal{H}^*\left(x^t, \mathbf{u}_\mathcal{R}\right) &= C_\mathcal{H}\left(x^t, \mathbf{u}_\mathcal{R}, g\left(x^t, \mathbf{u}_\mathcal{R}, \mathbf{u}_\mathcal{H}\right)\right)\end{aligned} \quad (5)$$

where $g\left(x^t, \mathbf{u}_\mathcal{R}, \mathbf{u}_\mathcal{H}\right)$ represents the response of the human driver towards the actions of the AV.

With the best response of the human driver for each possible action of the AV, the AV can find its best decision:

$$\mathbf{u}_\mathcal{R}^* = \arg\min_{\mathbf{u}_\mathcal{R}} C_\mathcal{R}\left(x^t, \mathbf{u}_\mathcal{R}, \mathbf{u}_\mathcal{H}^*\left(x^t, \mathbf{u}_\mathcal{R}\right)\right) \quad (6)$$

Based on the previously stated assumptions and the game formulation, we aim to generate altruistic behaviors of the AV by incorporating the courtesy factor in the AV's motion planning. In the following section, we will provide the implementation details of the socially compatible behavior planning of the AV.

## III. SOCIALLY COMPATIBLE BEHAVIOR PLANNING

A social factor such as altruism towards the human driver should be quantified and formulated as an additional feature into the cost function of the AV to achieve socially compatible behavior planning. The degree of the AV's altruism towards the human driver is defined as Social Value Orientation (SVO), a commonly used concept in the social psychology literature that has been recently integrated into robotics research [7-8]. We adopt the angular annotation for SVO in the socially compatible behavior planning, as defined in [16]. The SVO angular annotation $\phi$ quantifies how an agent weights its own reward against the rewards of another agent in the traditional interactive two-agent system. Therefore, the AV's cost function can be formulated as:

$$C_{\mathcal{R}} = C_e(x^t, \mathbf{u}_{\mathcal{R}})\cos(\phi) + C_c(x^t, \mathbf{u}_{\mathcal{H}})\sin(\phi) \quad (7)$$

where $C_e$ is the cost function for an egoistic AV which cares about only its own utilities; $C_c$ defines the courtesy term of the AV to the human driver; SVO angle $\phi$ ($\phi \in [0, \pi/4]$) defines the altruism level of the AV towards the human driver and cosine and sine functions are used to compute the weights of the AV's cost function with a given SVO angle. The intuitive explanation of two extreme SVO angles is that the AV behaves as an *egoistic* agent with $\phi = 0$ by maximizing only its own outcome, whereas it behaves as a *prosocial* agent with $\phi = \pi/4$ by maximizing the benefits of both itself and the following human driver, as expressed in [7].

*A. Egoistic Term*

The egoistic term denotes the effort of the AV to achieve its own driving goal when following the preceding traffic on the road. We formulate the egoistic term of the AV with the constant time headway (CTH) car-following strategy. The CTH strategy has been widely used as a speed planning strategy for AVs. It aims to maintain a constant time gap between the AV and its preceding vehicle, which ensures the desired speed of the AV is proportional to the gap distance [17]. Therefore, the egoistic term can be formulated as:

$$C_e(x^t, \mathbf{u}_{\mathcal{R}}) = \sum_{k=0}^{N-1} r_e(x^{t,k}, u_{\mathcal{R}}^k), \quad r_e(x^{t,k}, u_{\mathcal{R}}^k) = (d_{CTH} - d_{\mathcal{R}}^k)^2 \quad (8)$$

$$d_{CTH} = d_s + v_{\mathcal{R}}^k \tau_{\mathcal{R}} \quad (9)$$

where $d_{CTH}$ is the desired gap distance of the AV with the CTH strategy; $d_s$ is the minimum car-following gap distance and $\tau_{\mathcal{R}}$ is the constant time headway.

*B. Courtesy Term*

In this study, we model courtesy as the effort of the AV to avoid interrupting the human driver's original plan in the longitudinal driving scenario. In the car-following scenarios where the human drivers follow a preceding vehicle, human drivers are constrained by the preceding vehicle's actions and this may result in socially undesirable outcomes that are seen as uncomfortable and overly defensive by the human drivers. We assume that human drivers generally consider the speed limit of the traffic when driving on the road [18] and achieving and cruising at the speed limit of the traffic can be treated as the driver's original plan when the driver is not constrained by any preceding vehicle [14]. Therefore, we formulate the courtesy term as the deviation of the human driver's speed from the speed limit of the traffic:

$$C_c(x^t, \mathbf{u}_{\mathcal{H}}) = \sum_{k=0}^{N-1} r_c(x^{t,k}, u_{\mathcal{H}}^k), \quad r_c(x^{t,k}, u_{\mathcal{H}}^k) = (v_L - v_{\mathcal{H}}^k)^2 \quad (10)$$

where $v_L$ represents the speed limit of the traffic.

*C. Human Driver Behavior Model*

In the human-AV interaction, we have assumed that the AV can access the cost function of the human driver $C_{\mathcal{H}}$ to compute the human response and courtesy term, a common assumption in the framework of human-AV interactions [7-11]. Then the cost function $C_{\mathcal{H}}$ can be recovered from the human driving data through an offline learning process. We collect demonstrations of a driver in the driver-in-the-loop simulator [14] and use our previously developed inverse reinforcement learning (IRL) based driver behavior learning approach [13] to recover the cost function which explains the driver's driving preferences.

Based on the IRL approach, the cost function of the human driver is defined as a linear combination of the weighted features:

$$c_{\mathcal{H}}(x^t, u_{\mathcal{R}}^t, u_{\mathcal{H}}^t) = \mathbf{W}_{\mathcal{H}}^T \mathbf{f}_{\mathcal{H}}(x^t, u_{\mathcal{R}}^t, u_{\mathcal{H}}^t) \quad (11)$$

where $\mathbf{W}_{\mathcal{H}} = (W_1, W_2, \cdots, W_n)^T$ is the weight vector; $\mathbf{f}_{\mathcal{H}}(x^t, u_{\mathcal{R}}^t, u_{\mathcal{H}}^t) = (f_1, f_2, \cdots, f_n)^T$ is the feature vector and $n$ represents the number of defined features. The cost function over the planning horizon $N$ becomes:

$$C_{\mathcal{H}}(x^t, \mathbf{u}_{\mathcal{R}}, \mathbf{u}_{\mathcal{H}}) = \sum_{k=0}^{N-1} \left( \mathbf{W}_{\mathcal{H}}^T \mathbf{f}_{\mathcal{H}}(x^{t,k}, u_{\mathcal{R}}^k, u_{\mathcal{H}}^k) \right) \quad (12)$$

The goal is to find the weight vector $\mathbf{W}_{\mathcal{H}}$ that best describes the human demonstrations $\pi_{\mathcal{H}}^D \in \Pi_{\mathcal{H}}^D$ by maximizing the likelihood of the human driver behavior in the policy set $\Pi_{\mathcal{H}}^D$:

$$\mathbf{W}_{\mathcal{H}}^* = \arg\max_{\mathbf{W}_{\mathcal{H}}} P(\pi_{\mathcal{H}}^D \mid \mathbf{W}_{\mathcal{H}}) \quad (13)$$

According to the principle of Maximum Entropy, we assume that the human drivers are exponentially more likely to select trajectories with a lower cost:

$$P(\pi_{\mathcal{H}}^D \mid \mathbf{W}_{\mathcal{H}}) = \exp(-C_{\mathcal{H}}(x^t, \mathbf{u}_{\mathcal{R}}, \mathbf{u}_{\mathcal{H}})) \quad (14)$$

The weight vector $\mathbf{W}_{\mathcal{H}}$ can be derived with the gradient of the optimization problem. For the detailed derivation of the weight vector $\mathbf{W}_{\mathcal{H}}$ and more details about the driver behavior learning process, the reader is referred to [13].

**Features:** The features that are capable of describing fundamental longitudinal driving behaviors [9-14] are utilized to represent key driving behavior properties:

- *Acceleration*: capturing the ride comfort in the longitudinal direction:

$$f_a = (a_{\mathcal{H}})^2 \quad (15)$$

- *Desired speed*: achieving and cruising at the speed limit of the traffic:

$$f_{ds} = (v_L - v_{\mathcal{H}})^2 \quad (16)$$

- *Relative speed*: following the preceding vehicle's speed and maintaining a constant gap distance:

$$f_{rs} = (v_{\mathcal{R}} - v_{\mathcal{H}})^2 \quad (17)$$

- *Relative distance*: maintaining the desired car-following gap distance $d_D$ with the CTH strategy:

$$f_{rd} = |d_D - d_{\mathcal{H}}|, \quad d_D = v_{\mathcal{H}} \tau_{\mathcal{H}} + d_s \quad (18)$$

where $\tau_{\mathcal{H}}$ is the observed minimum time headway of the human driver from the demonstrations.

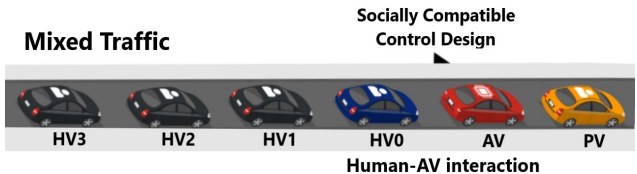

Fig. 1 Schematic of the socially compatible control design in mixed traffic.
*Source*: This figure was generated at https://icograms.com

**Trajectory Generation:** To generate driver-specific actions, a nonlinear MPC (NMPC) algorithm is used because of the nonlinear property of the recovered cost function. In the human-AV interaction, it is assumed that the human driver can correctly predict AV's motion in car-following scenarios if the preview time horizon is sufficiently short [10]. By using the definition of (12), the optimization problem of the human driver can be formulated as:

$$\mathbf{u}_{\mathcal{H}}^{*} = \arg\min_{\mathbf{u}_{\mathcal{H}}}\left(\mathbf{W}_{\mathcal{H}}^{T}\mathbf{f}_{\mathcal{H}}\left(x^{t},\mathbf{u}_{\mathcal{R}},\mathbf{u}_{\mathcal{H}}\right)\right), \mathbf{f}_{\mathcal{H}} = \left(f_{a},f_{ds},f_{rs},f_{rd}\right)^{T}$$
$$s.t.: d_{s} \leq d_{\mathcal{H}}^{k}, v_{\mathcal{H}_{min}} \leq v_{\mathcal{H}}^{k} \leq v_{\mathcal{H}_{max}} \quad (19)$$

where $v_{\mathcal{H}_{min}}$ and $v_{\mathcal{H}_{max}}$ are the minimum and maximum longitudinal speed constraints, respectively.

## IV. SOCIALLY COMPATIBLE CONTROL DESIGN IN MIXED TRAFFIC

In the previous section, we formulated the socially compatible behavior planning of the AV. Following that, we will describe the socially compatible control design for the AV to plan its longitudinal maneuvers in traffic. The proposed method minimizes the cost function (7) across the planning horizon by leveraging preview information from both the preceding vehicle (PV) and the following human driver (HV0). Such preview information can be acquired via vehicle connectivity and advanced sensing [3-4].

Furthermore, a homogenous human-driven fleet is introduced to simulate the following traffic for the human-AV interaction in the assessment of the proposed socially compatible control design's impacts on mixed traffic. The human-driven vehicles in the fleet are expressed as "HV1", "HV2" and "HV3" for clarity. Fig. 1 depicts the proposed design in mixed traffic.

### A. Control Design Implementation

In this section, an NMPC algorithm is formulated to solve the socially compatible control strategy of the AV. The control objective of the NMPC design is to compute the optimal control input vector of the AV $\mathbf{u}_{\mathcal{R}}^{*}$ for every time step $t$ by minimizing the cost function $C_{\mathcal{R}}$ within the prediction time horizon $N$ subject to the system constraints at each prediction step $k$. The minimum and maximum constraints are applied on the car-following gap distance of the AV $d_{\mathcal{R}}$ considering the safety and collision avoidance and allowable distance for vehicle connectivity, respectively. The constraints on the longitudinal speed $v_{\mathcal{R}}$ are applied by the determined minimum and maximum speed based on the traffic conditions. Finally, the constraints on the longitudinal acceleration $a_{\mathcal{R}}$ and control input $u_{\mathcal{R}}$ are used for the vehicle's drivability. The NMPC design parameters can be found in Table I. By applying (8) and (10), the optimal control problem of the AV can be formulated as:

$$\mathbf{u}_{\mathcal{R}}^{*} = \arg\min_{\mathbf{u}_{\mathcal{R}}}\left(C_{e}\left(x^{t},\mathbf{u}_{\mathcal{R}}\right)\cos(\phi)+C_{c}\left(x^{t},\mathbf{u}_{\mathcal{H}}\right)\sin(\phi)\right)$$
$$s.t.: d_{\mathcal{R}_{min}} \leq d_{\mathcal{R}}^{k} \leq d_{\mathcal{R}_{max}}, v_{\mathcal{R}_{min}} \leq v_{\mathcal{R}}^{k} \leq v_{\mathcal{R}_{max}} \quad (20)$$
$$u_{\mathcal{R}_{min}} \leq u_{\mathcal{R}}^{k} \leq u_{\mathcal{R}_{max}}, a_{\mathcal{R}_{min}} \leq a_{\mathcal{R}}^{k} \leq a_{\mathcal{R}_{max}}$$

### B. Microscopic Traffic Model

In this study, the Intelligent Driver Model (IDM) [19] is adopted as the microscopic traffic model to generate realistic traffic flow in mixed traffic, as shown in Fig. 1. This microscopic traffic model is considered as the following traffic for the human-AV interaction and consists of three human-driven vehicles. The model is expressed as follows:

$$\dot{v}=a\left(1-\left(\frac{v}{v_{s}}\right)^{\delta}-\left(\frac{s^{*}}{s}\right)^{2}\right), \quad s^{*}=s_{0}+\tau_{d}v+\frac{v\Delta v}{2\sqrt{ab}} \quad (21)$$

where $a$ is the maximum acceleration; $b$ is the comfortable deceleration; $v_{s}$ is the desired speed; $\Delta v$ is the speed difference of the subject vehicle to its preceding vehicle; $\tau_{d}$ is the desired time headway; $s^{*}$ is the desired gap distance; $s_{0}$ is the minimum gap distance and $\delta$ is the acceleration exponent. The model parameters are chosen from the realistic range of IDM parameters [20] and listed in Table I.

Table I: NMPC and IDM parameters.

| Parameter | Value | Parameter | Value |
|---|---|---|---|
| $N$ | 3 s | $a$ | 2 m/s² |
| $a_{\mathcal{R}_{min}}$ | -3 m/s² | $b$ | 2 m/s² |
| $a_{\mathcal{R}_{max}}$ | 3 m/s² | $s_0$ | 3 m |
| $u_{\mathcal{R}_{min}}$ | -4 m/s² | $\tau_d, \tau_{\mathcal{R}}$ | 1 s, 1.2 s |
| $u_{\mathcal{R}_{max}}$ | 4 m/s² | $\delta, \rho$ | 4, 0.45 |
| $d_{\mathcal{R}_{min}}, d_s$ | 5 m | $v_{\mathcal{H}_{max}}, v_{\mathcal{R}_{max}}, v_s$ | $\max(v_{PV})$ |
| $d_{\mathcal{R}_{max}}$ | 45 m | $v_{\mathcal{R}_{min}}, v_{\mathcal{H}_{min}}$ | 0 m/s |

## V. RESULTS AND DISCUSSIONS

By using the previously described models, we aim to investigate the impacts of the socially compatible control design on mixed traffic through different altruism levels of the AV. Therefore, we define four different altruism levels of the AV for the comparison study in which the SVO angles are described as $\phi \in [0, \pi/12, \pi/6, \pi/4]$. In this study, we performed an experiment to collect real-world driving data to verify the effectiveness of the proposed design with naturalistic vehicle trajectories. The driving scene consists of highway and urban driving scenarios, as shown in Fig. 2. This driving scene defines the PV's speed profile, and it is used to depict a realistic traffic scenario in which the PV's drivability is ensured. The speed profile can be found in Fig. 3 (PV).

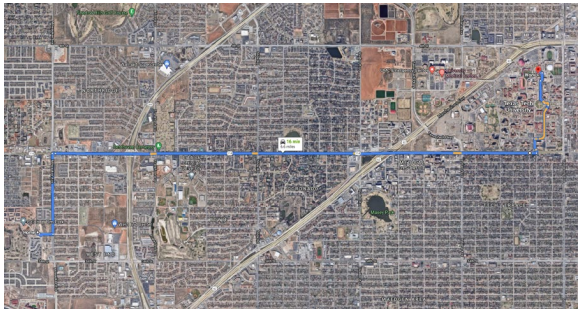

Fig. 2  Experimental daily commute driving scene.

We use the average gap distance and time headway of mixed traffic for each comparison case to assess the impacts of the socially compatible control design on mixed traffic. Each traffic participant's stated gap distance and time headway are calculated with respect to its preceding vehicle. We first evaluate the impacts of the socially compatible control strategy on the human-AV interaction with two extreme SVO angles. Fig. 3 and Fig. 4 show the speed and gap distance profiles of the traffic participants when the AV performs with egoistic $(\phi = 0)$ and prosocial $(\phi = \pi/4)$ behaviors, respectively. By comparing the gap distance profiles in Fig. 4, we find that the HV0 follows the prosocial AV quite closely than the egoistic AV. This is supported by Table II, where a 50-52% reduction in average gap distance and time headway of the HV0 is observed when the AV performs with prosocial behavior in the human-AV interaction. The fundamental reason for this is that the prosocial AV relieves the impedance towards the HV0 by incorporating the courtesy factor in its decision-making problem to avoid interrupting the HV0's original plan on the road.

Table II: Statistical comparison of the AV and HV0 in egoistic and prosocial altruism levels.

| Altruism Level | Average Gap Distance (AV-PV) | Average Gap Distance (HV0-AV) | Average Time Headway (AV-PV) | Average Time Headway (HV0-AV) |
|---|---|---|---|---|
| Egoistic | 23.28 m | 31.63 m | 1.62 s | 2.17 s |
| Prosocial | 19.91 m | 15.34 m | 1.29 s | 1.08 s |
| **Difference** | **14.45%** | **51.50%** | **20.55%** | **50.35%** |

We then evaluate the impacts of the socially compatible control strategy on mixed traffic. The results in Fig. 5 show the significant differences in the average gap distance and time headway of mixed traffic when comparing the traffic flow among the different altruism levels of the AV. It is found that the average gap distance and time headway of the traffic can be significantly reduced when the AV's altruism level increases toward prosocial. These results collectively demonstrate that the altruistic AV not only benefits the HV0 but also improves the traffic flow of mixed traffic with its increasing altruism level in human-AV interaction.

At last, we analyze the impacts of the socially compatible control design on mixed traffic by using a public driving dataset to demonstrate the effectiveness of the proposed design in various naturalistic vehicle trajectories. By this, we

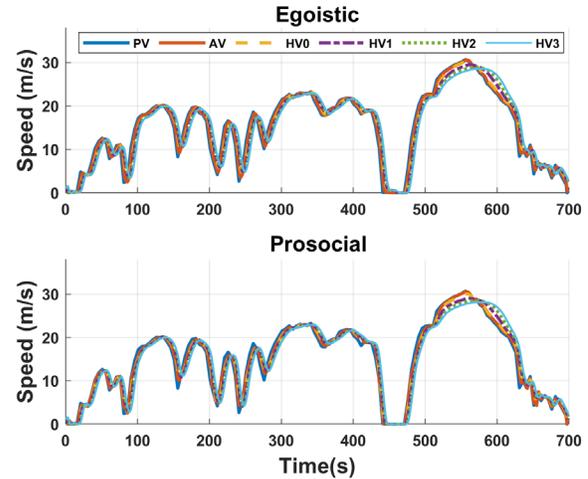

Fig. 3  Speed profiles comparison in egoistic and prosocial scenarios.

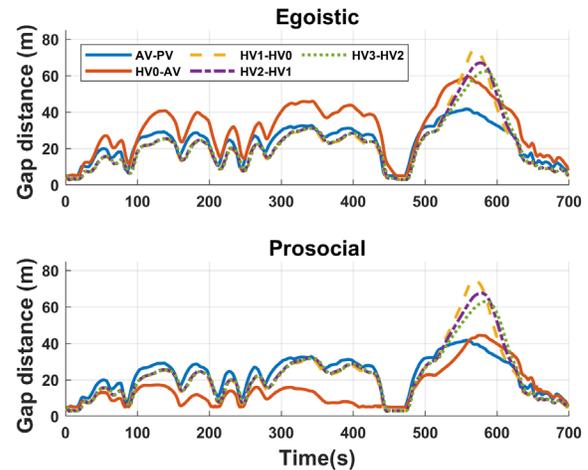

Fig. 4  Gap distance comparison in egoistic and prosocial scenarios.

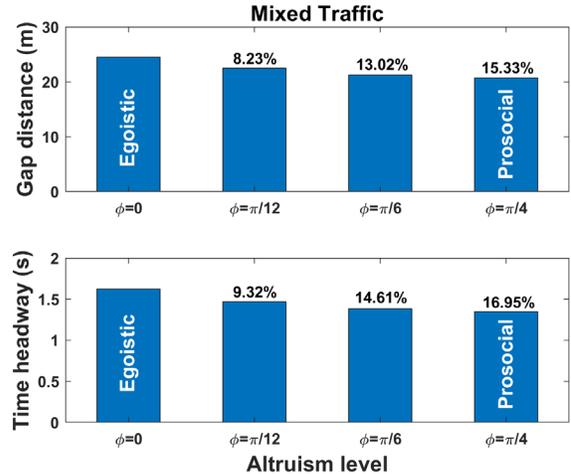

Fig. 5  Traffic average gap distance and time headway comparison among the different altruism levels of the AV.

randomly selected five different passenger vehicle trajectory data from the Next Generation SIMulation (NGSIM) I-80 dataset [21] and each vehicle's speed trajectory is assigned to PV's speed profile. The numerical results in Table III show that AV's prosocial behaviors provide a 3-6% decrease in average gap distance and a 4-8% decrease in average time headway of traffic compared to the egoistic behavior of the

AV in the human-AV interaction. These results indicate that the proposed socially compatible control design has the potential to improve the traffic flow of different realistic mixed traffic scenarios. Additionally, the benefits of the socially compatible control design on mixed traffic grow when the AV increases its altruism level toward the following human driver in the human-AV interactions.

Table III: Traffic flow difference when prosocial and egoistic AV participate in traffic (NGSIM I-80).

| Vehicle ID | Average Gap Distance | Average Time Headway |
|---|---|---|
| 70 | 4.24% | 6.27% |
| 17 | 4.23% | 6.04% |
| 182 | 3.76% | 5.54% |
| 25 | 5.85% | 8.12% |
| 291 | 3.84% | 4.49% |

## VI. CONCLUSION

**Summary:** In this work, we developed a socially compatible control design for the automated vehicle to create socially desirable outcomes that benefit itself and as well as the following human driver in the car-following scenarios. Furthermore, the impacts of the socially compatible control on mixed traffic are explicitly studied with simulation cases incorporating the automated vehicle's altruism variations in the human-AV interaction. The statistical results imply that the socially compatible behaviors of the automated vehicle can significantly improve the traffic flow of the mixed traffic, such as reducing the average gap distance and time headway.

**Limitations and Future Work:** Our work is a first step towards developing a socially compatible behavior for automated vehicles in the car following interactions with human drivers. We have so far assumed that the automated vehicle can acquire the cost function of the human driver through an offline learning process by using human demonstrations. In practice, automated vehicles may not access human demonstrations in advance to recover the driver-specific cost function offline and the offline learned cost function may mismatch with the behavior of the real driver. We also recognize that computing the Stackelberg game in the human-AV interaction can bring a high computational cost in real-time optimization. Based on these limitations, we will further extend the work by designing an online human driver behavior learning model in the human-AV interactions and developing real-time simulations to assess the performance of the proposed design.

Moreover, our preliminary results have shown that the human driver follows the altruistic automated vehicle with a considerably smaller gap distance and time headway compared with following the egoistic automated vehicle. Here, we argue that the altruistic automated vehicle with the proposed approach can potentially earn the trust of the following human driver and provide more comfortable car-following experiences, resulting in a smaller driver-perceived safety clearance. Hence, our future work will also focus on examining these hypotheses with proper and extensive statistical investigations.